\documentclass[aps,preprint]{revtex4}%
\usepackage{amsfonts}
\usepackage{amsmath}
\usepackage{amssymb}
\usepackage{graphicx}%
\begin{document}
\title{ \textbf{Thermodynamic Properties of electrically modulated monolayer
\ \ \ \ \ Graphene: Theory } }
\author{R. Nasir, M. A. Khan, M Tahir$^{\ast}$ and K. Sabeeh}
\affiliation{Department of Physics,Quaid-i-Azam University, Islamabad $45320$ Pakistan.}
\affiliation{$^{\ast}$Department of Physics, University of Sargodha, Sargodha $40100$, Pakistan.}
\keywords{one two three}
\pacs{PACS number}

\begin{abstract}
Theoretical investigation of thermodynamic properties of electrically
modulated monolayer graphene in the presence of a perpendicular magnetic field
$B$ is presented$.$ The results obtained are compared with those of the
conventional 2DEG. The one-dimentional periodic potential due to electric
modulation lifts the degeneracy of the Landau Levels and converts them into
bands whose width oscillates as the function of $B$. We find Weiss type
oscillations for small values of $B$ and dHvA type oscillations at larger
values values of $B$. These oscillations are more pronounced in Graphene than
in conventional 2DEG system. These oscillations are less damped with
temperature in Graphene compared with conventional 2DEG systems.

\end{abstract}
\volumeyear{year}
\volumenumber{number}
\issuenumber{number}
\eid{identifier}
\date[Date text]{date}
\received[Received text]{date}

\revised[Revised text]{date}

\accepted[Accepted text]{date}

\published[Published text]{date}

\startpage{1}
\endpage{2}
\maketitle

\section{\textbf{INTRODUCTION }}

Graphene is a 2D-honeycomb lattice of carbon atoms. Its experimental
realization has opened up new \ horizons in the field of condensed matter
physics and material sciences. Unique electronic properties of Graphene make
it substantially different from conventional 2DEG systems. The quasi particles
in graphene at low energies have a linear dispersion relation $\epsilon
_{k}=\hslash v_{F}k$ with characteristic velocity of $v_{F}=10^{6}m/s$%
\cite{1}.These quasi particles called massless Dirac Fermions, can be treated
as electrons with zero mass or neutrinos having electronic charge. The zero
mass property of charge carriers in graphene along with charge conjugation
symmetry, results in many unusual transport phenomena such as anomalous
Quantum Hall Effect, Chiral Tunneling and non-zero Berry's Phase\cite{2, 3, 4,
5, 6}. The 2D Dirac like spectrum was also confirmed recently by cyclotron
resonance measurements in monolayer Graphene\cite{1} and also by angle
resolved photo electron spectroscopy\cite{7}.

Weiss oscillations\cite{8, 9} appear in magnetoresistance when convential 2DEG
is subjected to artificially created periodic potentials (either electric or
magnetic) in submicron range. Electrical modulation can be carried out by
depositing an array of parallel metallic strips on the surface\cite{12} or
through two interfering laser beams\cite{13}. These Oscillations are the
direct consequence of the commensurability of two different length scales
namely the cyclotron orbit radius $R_{c}=\sqrt{2\pi n_{e}}l^{2}$ (where
$n_{e}$ is the density of electrons, and $l=\sqrt{\hslash/eB}$ is the magnetic
length) and the period of modulation $a$. Weiss oscillations occur in the
small magnetic field range\cite{10, 11} and are separate from dHvA(de Hass van
Alphen) and SdH(Subnikov de Hass) type oscillations which occur at larger
values of magnetic field. These oscillations are found to be periodic in the
inverse magnetic field. It is interesting to study the effects of electrical
modulation on Dirac electrons in graphene. Theoretical studies of tranport
properties of Dirac electron in graphene subjected to electrical modulation
were recently carried out and showed the appearance of Weiss oscillations in
magnetoconductivity. In addition, the magnetoplasmon spectrum of monolayer
graphene in the presence of electrical modulation was recently
investigated\cite{14}.

In this work we investigate the effects of elecrical modulation on
thermodynamic properties of monolayer graphene and compare the results
obtained with those of conventional 2DEG system found in semiconductor
hetrostructures. To this end, wehave determined the following thermodynamic
quantities:The chemical potential, Helmholtz free energy, orbital
magnetization , orbital magnetic susceptibility and electronic specific heat.
The results are compared with those of the conventional 2DEG studied
in\cite{15} and\cite{16}.

This paper is arranged as follows. In section II, we give the formulation of
the problem. The calculation of the thermodynamic quantities is given in
section III and numerical results with discussion are presented in section IV.
Finally the Conclusions are drawn at the end.

\section{\ FORMULATION}

We consider monolayer graphene in the $xy-plane$ subjected to a magnetic field
$B$ along the z-direction. In the Landau guage, the unperturbed Dirac like
Hamiltonian for single electron may be written as\cite{6}
\begin{subequations}
\begin{equation}
H_{o}=v_{F}\mathbf{\sigma}.\left(  -i\hslash\mathbf{\nabla}+e\mathbf{A}%
\right)  . \label{1}%
\end{equation}
Here, $\mathbf{\sigma}=\left\{  \sigma_{x},\sigma_{y}\right\}  $ are the Pauli
matrices and $v_{F}=10^{6}m/s$ characterizes the electron velocity. and
$\mathbf{A}=(0,Bx,0)$ is the vector potential.The normalized eigenfunctions of
the Hamiltonian given in Eq.(1)\cite{14 , 17}%
\end{subequations}
\begin{equation}
\Psi_{n,k_{y}}=\frac{e^{ik_{y}y}}{\sqrt{2L_{y}l}}\binom{-i\phi_{n-1}\left[
(x+x_{o})/l\right]  }{\phi_{n}\left[  (x+x_{o})/l\right]  }, \label{2}%
\end{equation}
\bigskip where $\phi_{n}=\frac{\exp(-x^{2}/2)}{\sqrt{2^{n}n!\sqrt{\pi}}}%
H_{n}(x)$, $H_{n}(x)$ are the Hermite Polynomials, $L_{y}$ is the
normalization length in the $y$-direction$,$ $n$ is an integer corresponding
to the Landau Level index and $x_{o}=k_{y}l^{2},$ is the center of the
cyclotron orbit. The energy eigenvalues are%
\begin{equation}
\varepsilon_{n}=\frac{v_{F}\hslash\sqrt{2n}}{l}=\sqrt{n}\hslash\omega_{c}
\label{3}%
\end{equation}
where $\omega_{c}=v_{F}\sqrt{\frac{2eB}{\hslash}}$ is the cyclotron frequency
of the Dirac electrons in graphene. To investigate the effects of modulation
we write the Hamiltonian in the presence of modulation as%
\begin{equation}
H=H_{o}+U(x) \label{4}%
\end{equation}
where $U(x)$ is the one-dimensional periodic modulation potential along the
$x$-axis and is given by
\begin{equation}
U(x)=V_{o}\cos Kx. \label{5}%
\end{equation}
$K=\frac{2\pi}{a}$, $a$ is the period of modulation and $V_{o}$ is the
constant modulation amplitude. To account for the weak modulation we take
$V_{o}$ to be an order of magnitude smaller than the Fermi Energy
$\varepsilon_{F}^{o}=v_{F}\hslash k_{F},$where $k_{F}=\sqrt{2\pi n_{s}}$ is
the magnitude of Fermi wave vector. Hence we can apply standard first order
perturbation theory to determine the energy eigenvalues in the presence of
modulation. The first order energy correction is%
\begin{equation}
\varepsilon_{n,x_{o}}=\varepsilon_{n}+U_{n}\cos Kx_{o} \label{6}%
\end{equation}
Here, $\ U_{n}=\frac{V_{o}}{2}\exp(-\frac{\chi}{2})[L_{n}(\chi)+L_{n-1}%
(\chi)]$, $\chi=\frac{K^{2}l^{2}}{2}$ and, $L_{n}(\chi)$ and $L_{n-1}(\chi
)\ $are Laguerre polynomials.

Although similar features in the energy spectrum have also been found in the
2DEG system under similar conditions,\cite{15 , 16} there are substantial
differences between the two systems. Landau level spectrum of Dirac electrons
depends on the square root of both magnetic field $B$ and the Landau band
index $n$ against linear dependence in the case of conventional electronsin
2DEG. The energy eigenvalues in the presence of modulation given by Eq.(6)
contains a term which is a linear combination of\ two succesive Laguerre
polynomials with indices $n$ and $n-1$ , while conventional electrons obey a
relation containg one Laguerre polynomial with index $n$.

The modulation potential lifts the degeneracy of the Landau levels and
broadens the formerly sharpe levels into electric Landau bands. The electric
modulation induced broadening of the energy spectrum is nonuniform. The Landau
band width $U_{n}$ oscillates as a function of $n$ since $L_{n}(\chi)$ is
anoscillatory function of the index $n$. These landau bands become flat for
different values of $B$. Flat bands occure for those values of $B$ for which
modulation strength becomes zero. By putting $Un=0$ one can get the flat band
condition.%
\begin{equation}
\exp(-\frac{\chi}{2})[L_{n}(\chi)+L_{n-1}(\chi)]=0 \label{7}%
\end{equation}
using the asymptotic expression\cite{17}%
\begin{equation}
\exp(-\frac{\chi}{2})L_{n}(\chi)\simeq\frac{1}{\sqrt{\pi\sqrt{n\chi}}}%
\cos(2\sqrt{n\chi}-\frac{\pi}{4}) \label{8}%
\end{equation}
and $L_{n}(\chi)=L_{n-1}(\chi),$ one obtains from Eqs (6) and (7) the
following condition%
\begin{equation}
2R_{c}=a(i-1/4),\text{\ \ \ }i=1,2,3,.......... \label{9}%
\end{equation}
where, $R_{c}=k_{F}l^{2},$ is the classical cyclotron orbit. From Eqs $(6)$
and $(8)$ it can be observed that, in the large $n$ limit electron bandwidth
oscillates sinosoidally and is periodic in $1/B,$ for fixed values of $n$ and
$a.$When $n$ is small bandwidth still oscillates, but the condition $(9)$ no
longer holds because neigther eq. $(8)$ nor $L_{n}(\chi)\simeq L_{n-1}(\chi)$
is valid. Interestingly, for low values of $B$, when many Landau levels are
filled, both the systems have the same flat band condition\cite{15}.

It is well known that in the absence of modulation the density of states (DOS)
consists of a series of delta functions at energies equal to $\varepsilon_{n}%
$. The addition of a weak periodic electric modulation however modifies the
former delta functions leading to DOS broadening . The density of states
$D(\varepsilon)$ are given by \cite{18}%
\begin{equation}
D(\varepsilon)=\frac{A}{\pi l^{2}}\underset{n,x_{o}}{%
{\displaystyle\sum}
}\delta\left(  \varepsilon-\varepsilon_{n,x_{o}}\right)  =\frac{A}{\pi l^{2}%
}\underset{n,x_{o}}{%
{\displaystyle\sum}
}\frac{\ \theta\left(  \left\vert U_{n}\right\vert -\left\vert \varepsilon
-\varepsilon_{n,x_{o}}\right\vert \right)  }{\sqrt{\left\vert U_{n}\right\vert
^{2}-\left(  \varepsilon-\varepsilon_{n,x_{o}}\right)  ^{2}}} \label{10}%
\end{equation}
where, $\theta(x)$ is a unit Heaviside step function and $A$ is the area of
the sample.

\section{\textbf{EQUILIBRIUM\ THERMODYNAMIC\ QUANTITIES}}

We have determined the electronic contribution to the equillibrium
thermodynamic properties of monolayer graphene subjected to a perpendicular
magnetic field and weak electric modulation. The thermodynamic quantities
calculated are chemical potential, Helmholtz free energy, electronic specific
heat, orbital magnetization and orbital magnetic susceptibility.

The magnetid field ($B$) and temperature ($T$) dependent chemical potential
$\mu\equiv\mu(B,T)$ of a system can be determined by inverting the following
relation%
\begin{equation}
N=\overset{\infty}{\underset{0}{%
{\displaystyle\int}
}}D(\varepsilon)f(\varepsilon)d\varepsilon\label{11}%
\end{equation}
where the Fermi Dirac distribution function $f\left(  \varepsilon\right)  $
is
\begin{equation}
f\left(  \varepsilon\right)  =\left[  \exp\left(  \frac{\varepsilon-\mu}%
{k_{B}T}\right)  +1\right]  ^{-1},\label{12}%
\end{equation}
$k_{B}$ is the Boltzmann's constant and $N$ is the total number of electrons.
Hence change in the $D(\varepsilon)$ will affect $\mu(B,T)$. Substituting
Eq.(9) into Eq.(11) we obtain%
\begin{equation}
N=\frac{A}{\pi^{2}l^{2}}\overset{\infty}{\underset{n=0}{\sum}}%
{\displaystyle\int\limits_{-1}^{1}}
\frac{dx}{\sqrt{1-x^{2}}}(1+\chi_{n}\exp[z_{n}x])^{-1}\label{13}%
\end{equation}
Here $\chi_{n}=\exp\left[  \frac{\varepsilon_{n}-\mu}{k_{B}T}\right]  $ and
$z_{n}=\left\vert U_{n}\right\vert /(k_{B}T).$ Eq.(12) can be used for both
modulated and unmodulated systems ($z_{n}=0)$. We solve this equation
numerically in order to obtain the chemical potential $\mu(B,T)$. We are able
to determine the change in the chemical potential due to the electric
modulation. Once the chemical potential and the density of states are known,
the free energy $F$ of the system can be calculated. From there on the
thermodynamic properties of the system can be obtained from the free energy by
taking the appropriate derivatives. For a system of non-interacting fermions,
the Helmholtz free energy is given by \cite{19}
\begin{equation}
F=\mu N-k_{B}T\overset{\infty}{\underset{0}{%
{\displaystyle\int}
}}D(\varepsilon)\ln\left[  1+\exp\left(  \frac{\mu-\varepsilon}{k_{B}%
T}\right)  \right]  d\varepsilon\label{14}%
\end{equation}
The density of states $D(\varepsilon)$ is the central quantity in the above
expression. The expression for $D(\varepsilon)$ in graphene is different from
that in conventional 2DEG due to the difference in the energy spectrum in the
two cases. This difference will affect the electronic contribution in the
thermodynamic properties in the two systems determined from the following free
energy for the electrically modulated graphene system
\begin{equation}
F=\mu N-k_{B}T\text{ }\frac{A}{\pi^{2}l^{2}}\overset{\infty}{\underset
{n=0}{\sum}}%
{\displaystyle\int\limits_{-1}^{1}}
\frac{dx}{\sqrt{1-x^{2}}}\ln\left[  1+\chi_{n}^{-1}\exp\left(  -z_{n}x\right)
\right]  \label{15}%
\end{equation}

From Eq. 15, one can calculate the electronic comtribution of magnitization
for both graphen and 2DEG systems as $M=-\left(  \frac{\partial F}{\partial
B}\right)  _{A,\;N}$ and specific heat as $C_{v}=-T\left(  \frac{\partial
^{2}F}{\partial T^{2}}\right)  _{A,N}.$ The electronic contribution to
susceptibility is obtained directly from $\chi=-\left(  \partial^{2}F/\partial
B^{2}\right)  .$

\section{RESULTS AND DISCUSSION}

Numerical study of thermodynamic properties for monolayer graphene system
subjected to electrical modulation is presented. We have also plotted the same
quantities for the 2DEG system \ This is to facilitate comparison and was also
a check on our numerical program. For the 2DEG parameters for GaAs are used.
We have taken $n_{s}=3.16\times10^{15}m^{-2}$ and $a=382nm$. For electrical
modulation we have taken $V_{0}=1meV$. Thus our 2DEG results are those already
given in\cite{15 , 16}. Modulation induced effects on thermodynamic quantities
can be highlighted by calculating the difference between the modulated case
and the unmodulated case in each system.

In Figures $1-5$ we have plotted the change in various thermodynamic
properties due to electric potential at temperatures of $T=2K$ (full curve)
and $T=6K$ (broken curve).for both conventional 2DEG system and graphene
system. These figures were scaled to approperiate values to make them appear dimensionless.

In Fig.(1), we have plotted the change in chemical potential versus magnetic
field at temperatures $2K$(straight) and $6K$(broken). For Conventional 2DEG
system for $B<0.3T$ oscillations depend very weakly on temperature, which is a
clear signature of Weiss type Oscillations. Where as for $B>0.3T$, the
oscillations depends strongly on temperature, in particular they die out at
$6K$, a clear signature of dHvA type oscillations. Furthermore, the zeros in
the chemical potential are in close agreement as predicted by the flat band
condition Eq.(9). A similar behavior is expected for Graphene system. But for
Graphene the value of $B$ defining the boundary between the two oscillatory
phenomena is quite low (It lies some where between $0.1$and $0.15T$). For
smaller values of $B$ Weiss type oscillations are present and the amplitude of
the oscillations remain essentially the same at different temperatures. For
larger values of $B,$ the familiar dHvA-type oscillations are present ,as the
amplitude of oscillations is reduced considerably at comparatively higher
temperature. However the oscillatory phenomenon still persists, contrary to
the conventional 2D system in which oscillations completely die out at $6K$.
In comparison we can say, Graphene system is more sensitive to the magnetic
field and less sensitive to temperature, than the conventional 2DEG system.
This difference arises mainly due to the difference in the Landau level
energies of the two systems and due to the presence of an additional Laguerre
Polynomial term in the modulation contribution to the energy spectrum for
Graphene system

The Free energy is shown in Fig.(2), for the two systems. To make $y$-axis
dimensionless, Free energy has been scaled using $F_{0}=%
\frac12
NE_{F}$. It can be seen that at small values of $B$ periodically modulated
potential induces temperature independent Weiss type oscillations, with zeros
occurring at their respective flat band conditions. Weiss Oscillations are
more prounounced in Graphene system, significantly the amplitude of Weiss
oscillations for the graphene system remains unchanged at higher temperature,
contrary to the 2DEG in which damping may be observed. The familiar dHvA type
oscillations are observed for higher values of $B$. As in the case of the
Chemical Potential, again the dHvA type oscillations starts quite early.The
first period for the dHvA type oscillations starts from $B=0.3T$ and extends
up to $0.6T$ \ for the standard 2DEG system, while for graphene the first
period of dHvA type starts near $B=0.175T$ and terminates at{} $0.27T.$

In Figs.(3) and (4) we have plotted the changes in the magnetization $\Delta
M$ and the susceptibility $\Delta\chi$ against the magnetic field. Both the
quantities has been approperiately scaled to appear dimensionless. At low $B$
oscillations having their origin in the commensurability of two length
scales,and are effected weakly by temperature, having zeros as given by their
respective flat band conditions. For higher values of $B,$ dHvA oscillations
are present at lower temprature $(2K),$ with amplitude becomeing zero for the
conventional 2DEG system while reduced considerably for the Graphene system at
higher temprature $(6K)$.

In fig.5 we plot change in the electronic specific heat capacity against
magnetic field. $y$-axis \ has been scaled using $C_{el}=Nk_{B}$, to appear
dimensionless. In both systems the Weiss type oscillations are not large
effects, however the damping behavior of dHvA type oscillations is clearly observeable.

\section{CONCLUSIONS}

We have presented a study of the thermodynamic properties of monolayer
graphene system and compared the results with those of the conventional 2DEG.
The commensurability oscillations(Weiss type) and dHvA type oscillations are
reflected in all the thermodynamic quantities under consideration in this work
for the two systems. However, these effects are more prounounced in graphene
system in the sense that the oscillations in the thermodynamic quantities are
more robust against temperature. We can therefore say that Graphene system is
less sensitive to temprature and more sensitive to the magnetic field. This
differnce arises because of the different nature of the quasiparticles in the
two systems.

\end{document}